\documentclass[11pt]{article}
\usepackage[utf8]{inputenc}
\usepackage[margin=1in]{geometry}
\usepackage{enumerate}
\usepackage{amsmath,amssymb}
\usepackage{algpseudocode}
\usepackage{mathtools}
\usepackage{csquotes}
\usepackage{mathtools}
\mathtoolsset{centercolon}
\usepackage{color}

\newcommand{\hugeDebug}{false}
\ifthenelse{\equal{\hugeDebug}{false}}{
\newcommand{\normalspacing}{\singlespacing}
\pagestyle{plain}

\setlength{\oddsidemargin}{0.25in}
\setlength{\evensidemargin}{\oddsidemargin}
\setlength{\textwidth}{6in}
\setlength{\textheight}{8in}
\setlength{\topmargin}{-0.0in}
}
{

\newcommand{\normalspacing}{\niceninespacing}
\pagestyle{headings}
 \setlength{\oddsidemargin}{-0.4in}
 \setlength{\evensidemargin}{\oddsidemargin}
 \setlength{\textwidth}{5.3in}
 \setlength{\textheight}{6in}
 \setlength{\textheight}{6.25in}
 \setlength{\topmargin}{-0.0in}
}

\makeatletter
\newcommand{\singlespacing}{\let\CS=
\@currsize\renewcommand{\baselinestretch}{1}\tiny\CS}
\newcommand{\niceninespacing}{\let\CS=\@currsize\renewcommand{\baselinestretch}{1.9}\tiny\CS}
\makeatother

\normalspacing

\makeatletter
\g@addto@macro\bfseries{\boldmath}
\makeatother
\newcommand{\stringset}{\ensuremath{\{0,1\}^*}}
\newcommand{\pz}{\ensuremath{\hat{\rm P}}}
\newcommand{\npz}{\ensuremath{\hat{\rm NP}}}

\title{Critique of Barbosa's ``P != NP Proof"}
\author{Jackson Abascal\thanks{
Work supported in part by a CRA-W CREU grant.
}~~and Shir Maimon\footnotemark[1]
\\Department of Computer Science \\
University of Rochester \\
Rochester, NY 14627, USA
}

\begin{document}
\sloppy

\maketitle

\begin{abstract}
We review Andr\'e Luiz Barbosa's paper ``P != NP Proof," in which the classes P and NP are generalized and claimed to be proven separate.
We highlight inherent ambiguities in Barbosa's definitions, and show that attempts to resolve this ambiguity lead to flaws in the proof of his main result.

\end{abstract}

\section{Introduction}
Despite its provocative title, Andr\'e Luiz Barbosa's paper ``P != NP Proof"\footnote{This critique is written with respect to the most recent available revised version: Version 77 (the version of November 6, 2016) of arXiv.org report 0907.3965~\cite{bar:t:p-neq-np}.} 
does not claim to separate P from NP.
Instead, the paper provides ``generalized" definitions of P and NP, as well as a claimed proof that these generalized classes are not equal.
In particular, he defines a language XG-SAT and gives an argument that XG-SAT is in the difference of these classes.
We will show that attempts to resolve vagueness in Barbosa's definitions lead to consequences ignored in the paper, including the failure of his separation proof for redefined P and redefined NP.

\section{Preliminaries}
We will reiterate some of Barbosa's definitions.
Let $L_z \subset \stringset$ (the subscript $z$ in $L_z$ is not an indexing variable, but instead a distinguishing mark).
Barbosa calls a subset $L \subseteq L_z$ an $L_z$-language.
Throughout the paper, he uses the ambient set $L_z$ of an $L_z$-language $L$ as if $L$ somehow contains an underlying reference to $L_z$.
Hereafter, when we refer to an ``$L_z$-language $L$" it is assumed that $L$ and $L_z$ are linked to one another, and that what we call $L$ is implicitly an ordered pair $(L_z,L)$.
A machine is said to decide an $L_z$-language $L$ if it accepts on all $w \in L$ and rejects on all $w \in L_z-L$.
Any machine behavior is acceptable on strings in $\stringset-L_z$, including running forever.
Let $S$ be a deterministic machine.
We call $S$ a restricted type X program if there is a polynomial $P$ such that for all $n > 0$ one of the following holds:
\begin{enumerate}
\item[(1)] $S$ rejects every input of length $n$, each in at most $P(n)$ steps or
\item[(2)] $S$ accepts some input of length $n$ in at most $P(n)$ steps.
\end{enumerate}
Barbosa's separation argument revolves around a particular $L_z$-language called XG-SAT.
Here, $L_z$ is the set of strings of the form $1^n0\langle S \rangle $ where $n > 0$ and $\langle S \rangle$ is the encoding of a restricted type X program $S$.
A string $1^n0\langle S \rangle$ is in XG-SAT if $S$ accepts at least one string of length $n$.

Barbosa redefines the concept of a deterministic polynomial time machine.
He uses the phrase ``Poly-time DTM" to refer to his definition, and we will do so as well.
Barbosa's exact definition of a Poly-time DTM in the paper is as follows:
\begin{quote}
\textbf{Poly-time DTM.}
A DTM is said to be polynomial-time if its running time
T(n) = $O$(n\textsuperscript{k}), for some finite
nonnegative \textbf{k} that \underline{does not depend on \textbf{n}}. (n = $|$input$|$.) \cite{bar:t:p-neq-np}
\end{quote}

Referring to this definition, Barbosa says,
``the polynomial CAN definitely depend on the input - as long as it does not depend on the input’s length."
This implies that $k$ is not necessarily a constant but instead a function of the input.
This raises some immediate concerns.
First, Big O notation describes functions with numerical inputs.
If $k$ is a function with domain $\stringset$, then $n^k$ is also a function with domain $\stringset$, namely the function sending $w$ to $|w|^{k(w)}$.
We can naturally extend the definition of Big O notation to include these functions: For two functions $f,g:\stringset \to \mathbb{N}$ we say that $f$ is $O(g)$ if there exists a constant $c$ such that $f(w) \leq c \cdot g(w)$ for all but finitely many $w$.
Another glaring concern is the vague condition that $k$ ``does not depend" on the length of the input.
Given that $k$ is explicitly allowed to depend on the input itself, a precise definition of independence is needed.
None is given in the paper.
We present several natural definitions of independence and show that Barbosa's separation proof fails under each of them.

\begin{enumerate}
\item[(1)] Given functions $f,g:\stringset\to\mathbb{N}$, we say that $f$ is independent of $g$ if there does not exist $h:\mathbb{N}\to\mathbb{N}$ such that $f = h \circ g$.

\item[(2)] Given functions $f,g:\stringset\to\mathbb{N}$, we say that $f$ is independent of $g$ if for all $k$, $f(\stringset) = f(g^{-1}(k))$ for all but finitely many $k$.

\item[(3)] Given functions $f,g:\stringset\to\mathbb{N}$, we say that $f$ and $g$ are independent if there exists any probability distribution $\mu$ on $\stringset$ such that $f$ and $g$ are independent random variables under $\mu$, and $\mu(\{x\})>0$ for all but finitely many $x \in \stringset$.\footnote{One might ask why we impose the restriction that $\mu(\{x\})>0$ for all but finitely many $x \in \stringset$.
If this restriction is removed, every two functions $f,g:\stringset\to\mathbb{N}$ are independent.
Let $\mu(\{0\}) = 1$ and $\mu(\{x\})=0$ for all $x \neq 0$.
Then
$P(f \in A) \cdot P(g \in B)=1$ if and only if $f(0) \in A$ and $g(0) \in B$, and $P(f \in A) \cdot P(g \in B)=0$ otherwise.
Similarly, $P(f\in A$ and $g \in B)=1$ if and only if $f(0) \in A$ and $g(0) \in B$, and $P(f\in A$ and $g \in B)=0$ otherwise.
Thus, $f$ and $g$ are independent.}
\end{enumerate}

Definition (1) formalizes the idea that for $f$ to be independent of $g$, it cannot be written as a function of the output of $g$.
Given this context, to say that $k(w)$ is independent of $|w|$ means that $k(w)$ is not solely a function of the length of $w$.
Let $M$ be a machine deciding an $L_z$-language $L$.
For $w \in L_z$, define $s(w)$ as the number of steps $M$ takes on input $w$ before halting, and define $r(w)$ as the number of 1's in $w$.
Let $k(w) = 2s(w) + r(w)$.
For $|w| > 1$,
\[s(w) \leq |w|^{s(w)} \leq |w|^{2s(w) + r(w)} = |w|^{k(w)} \]
so $M$ will always halt on input $w$ in time $O(|w|^{k(w)})$.
Furthermore, $k$ is not solely a function of $|w|$, since for $n > 1$, $k(0^{n})$ and $k(0^{n-1}1)$ have different parities and thus cannot be equal, despite the fact that $0^n$ and $0^{n-1}1$ have the same length.
Thus under Definition (1), every decider is a Poly-time DTM.

Definition (2) formalizes the idea that knowing the value of $g(w)$ almost never gives information about the value of $f(w)$.
Given this context, it means that knowing $|w|$ gives no information about $k(w)$ for all but finitely many possible values of $|w|$.
Let $\ell(w) = |w|$, and suppose $k$ is independent of $\ell$.
Let $N$ be the minimum value such that for all $n \geq N$, $k(\stringset) = k(\ell^{-1}(n))$.
There are only finitely many strings of length $N$, so $k(\ell^{-1}(n))$ is finite and $k(\ell^{-1}(n))$ is bounded by its maximum value.
Then $k(\stringset)$ is also finite and $k$ is bounded.
Therefore, if a Poly-time DTM runs in time $O(|w|^{k(w)})$, it also runs in time $O(|w|^{k'})$ for some fixed constant $k'$ bounding $k$.
The definition of a Poly-time DTM in this case is equivalent to the traditional definition.

Definition (3) draws on the formalization of independence in probability theory.
Let $\ell(w) = |w|$, and suppose $\mu$ is a probability distribution on $\stringset$ such that $\mu(\{w\}) = 0$ for only finitely many $w \in \stringset$, and $k(w)$ and $\ell(w)$ are independent random variables.
Choose some $w \in \stringset$ such that $\mu(\{w\}) > 0$.
Let $n = \ell(w)$, and let $A = k(\ell^{-1}(n))$.
There are only $2^n$ elements of $\ell^{-1}(n)$, so $A$ is finite.
Since $k$ and $\ell$ are independent, they satisfy
\[P(k \in A ) \cdot P(\ell \in \{n\}) = P(k \in A \text{ and }\ell \in \{n\}).\]

If $\ell(w') = n$, then $k(w') \in A$ by definition so $P(k \in A \text{ and }\ell \in \{n\}) = P(\ell \in \{n\})$.
Then $P(k \in A ) \cdot P(\ell \in \{n\})=P(\ell \in \{n\})$ so either $P(k \in A )=1$ or $P(\ell \in \{n\})=0$.
But $P(\ell \in \{n\}) \geq \mu(\{w\})>0$ so $P(k \in A) = 1$.

The set $A$ is finite, and $P(k\in A) = 1$.
Thus for every $s \in \stringset$ with $\mu(\{s\}) > 0$, $k(s)$ is bounded by the maximum value in $A$.
Let $B=\{k'\,|\,k(s)=k'$  for some string $s$ such that $\mu(\{s\})=0\}$.
There are only finitely many strings with zero probability, so $B$ is finite; therefore $A \cup B$ is finite and has a maximum value $k'$.
For all strings $s$, $k(s) \leq k'$.
A Poly-time DTM of runtime $O(|w|^{k(w)})$ therefore also has runtime $O(|w|^{k'})$, so just as in Definition (2), a Poly-time DTM is exactly a traditional polynomial-time deterministic machine.

\section{P-versus-NP}

Barbosa redefines the classes P and NP.
To avoid confusion, we will refer to Barbosa's redefined P and NP as \pz{} and \npz{} respectively, although he does not use this notation in his paper.
An $L_z$-language $L$ is in the class \pz{} if there exists a Poly-time DTM which decides correctly for all $w \in L_z$ whether $w \in L$.
This $L_z$ is not the same for all languages in \pz{}.
A language is in \npz{} if there is a polynomial $p$ and a Poly-time DTM $M$, called a verifier, such that for all $w \in L_z$, $w \in L$ if and only if there exists a string $x$ with length at most $p(|w|)$ such that $M$ accepts $w\#x$.

\subsection{Complexity of XG-SAT}
Barbosa attempts to show that XG-SAT is in \npz{} by providing a verifier.
He states that the verifier should take a string of the form $1^n0\langle S \rangle\#x$ where $\langle S \rangle$ is the encoding of a restricted type X program $S$ and $|x| = n$, and accept if $S$ accepts $x$ ``within polynomial time." This phrase implies that after some polynomial number of steps simulating $S$ on $x$, the verifier should halt.
Such a bounding polynomial, however, is specific to every machine $S$ and not fixed for any verifier.
It is not clear how Barbosa intends to determine a maximum number of steps after which to cut off any computation, and so this description is not sufficient to construct a verifier for XG-SAT.

The format of Barbosa's flawed verifier construction does shed some light on the choice of definition of a Poly-time DTM.
Barbosa would like to be able to take as input arbitrary polynomial-time machines and simulate them, while keeping the simulator in polynomial time as well.
He attempts to do this by stating that the runtime of a machine may depend on the input but not its length while neglecting to consider the possible repercussions of this definition.

\subsection{Either \pz{} = \npz{} or XG-SAT is not in \npz{}}

Under Definition (1), \pz{} and \npz{} are equal.
Recall that Definition~(1) implies every decider is a Poly-time DTM.
Then \pz{} is the class of all decidable $L_z$-languages.
\npz{} is also the class of all decidable $L_z$-languages since any \npz{} language $L$ can be decided deterministically as follows: If $M$ is an \npz{} verifier for $L$, on input $w$ we will run $M$ on all strings of length at most $p(|w|)$ and accept if $M$ ever accepts.
In this case, \pz{}~$=$~\npz{}, so Barbosa cannot possibly separate the classes.

Under Definitions (2) and (3), XG-SAT is not in \npz{}.
The argument we give below is in fact adapted from page 9 of Barbosa's paper.
Suppose XG-SAT is in \npz{}.
Then there exists a verifier $T$ for XG-SAT and polynomial $p$ such that $T$ runs for at most $p(|w|)$ steps for any valid input $w$.
Valid inputs are of the form $1^i0\langle S \rangle\#x$ where $\langle S\rangle$ is the encoding of a restricted type X program.
In \cite{Seiferas:1978:SNT:322047.322061}, Seiferas, Fischer, and Meyer prove the existence of a decidable unary language $L$ which is not in NP (note that NP here refers to the classical complexity class, not \npz{}).
Let $M$ be a machine deciding $L$ and let $M_t$ be the machine which simulates $M$ on inputs of length $t$ and rejects all other inputs.
Clearly $M_t$ runs in constant time since it will only simulate $M$ for finitely many inputs and reject all other inputs after at most $t+1$ steps.
Therefore $M_t$ is a restricted type X program.
We can now create a decider $M'$ for $L$ which runs in nondeterministic polynomial time.
On input $w$, if $w$ is of the form $1^t$, $M'$ guesses a polynomial length string $x$ and simulates $T$ on input $1^t0\langle M_t\rangle\#x$, where $\langle M_t \rangle$ is the encoding of $M_t$.
Otherwise $M'$ rejects.
Therefore $L$ is in NP, which is a contradiction.

\paragraph*{Acknowledgments} We thank Michaela Houk, Daniel Rubery, and Jesse Stern for their comments on previous drafts of this paper.
We also thank Professor Lane A. Hemaspaandra for being our CREU mentor and presenting us with the original paper.

\bibliographystyle{alpha}
\bibliography{refs.bib}

\begin{thebibliography}{SFM78}

\bibitem[Bar11]{bar:t:p-neq-np}
A.~Barbosa.
\newblock {P} != {NP} proof.
\newblock Technical Report arXiv:0907.3965~[cs.CC], Computing Research
  Repository, \mbox{arXiv.org/corr/}, May 2011.
\newblock Version 77 (of November 6, 2016); the date of Version 1 was July 23,
  2009.

\bibitem[SFM78]{Seiferas:1978:SNT:322047.322061}
J.~Seiferas, M.~Fischer, and A.~Meyer.
\newblock Separating nondeterministic time complexity classes.
\newblock {\em J. ACM}, 25(1):146--167, January 1978.

\end{thebibliography}

\end{document}